%% file: main.tex
\documentclass[conference]{IEEEtran}

\usepackage{cite}
\usepackage[inline]{enumitem}

\usepackage{cuted}

\usepackage{booktabs} % for drawing tables lines spanning few cols

\usepackage{graphicx}
\usepackage[justification=centering]{caption}
\usepackage{wrapfig}
\usepackage{subcaption}

\usepackage{color}

\usepackage{amsmath}
\usepackage{multirow}
\usepackage{array}
\newcolumntype{P}[1]{>{\centering\arraybackslash}p{#1}}

\usepackage[table]{xcolor}
\definecolor{blue_t}{RGB}{221,221,255}
\newcommand{\bc}[1]{\cellcolor{blue_t}}

\definecolor{red_t}{RGB}{255,221,221}

\definecolor{green_t}{RGB}{169,208,142}
\newcommand{\gc}[1]{\cellcolor{green_t}}

\definecolor{peach_t}{RGB}{244,176,132}
\newcommand{\pc}[1]{\cellcolor{peach_t}}

\definecolor{grey_t}{RGB}{201,201,201}
\newcommand{\grc}[1]{\cellcolor{grey_t}}

\definecolor{yellow_t}{RGB}{255,192,0}
\newcommand{\yc}[1]{\cellcolor{yellow_t}}
\usepackage{nicematrix}

\begin{document}

\bstctlcite{IEEEexample:BSTcontrol}

\title{Design and Optimization of  Magnetic-core Solenoid  Inductor for Multi-phase Buck Converter \vspace{-0.5cm}}

\author{ \IEEEauthorblockN{Madhava Sarma Vemuri} \IEEEauthorblockA{ Department of Electrical and Computer Engineering\\
  North Dakota State University\\
  Fargo, North Dakota 58105 \\
  %\texttt{madhava.vemuri@ndsu.edu}
  }
\and
\IEEEauthorblockN{Umamaheswara Rao Tida} \IEEEauthorblockA{ Department of Electrical and Computer Engineering\\
  North Dakota State University\\
  Fargo, North Dakota 58105 \\
 %\texttt{umamaheswara.tida@ndsu.edu}
  }
\thanks{Madhava Sarma Vemuri is with the Department of Electrical and Computer Engineering, North Dakota State University, North Dakota, United States -- 58102}% <-this % stops a space
\thanks{Umamaheswara Rao Tida is with the Department of Electrical and Computer Engineering, North Dakota State University, North Dakota, United States -- 58102. e-mail: umamaheswara.tida@ndsu.edu}% <-this % stops a space
\thanks{Manuscript received: Oct 14, 2019;}
\vspace{-1cm}}% revised \redHL{Date.}}}

\title{Metal Inter-layer Via Keep-out-zone in M3D IC: A Critical
Process-aware Design Consideration (Accepted in ISQED 2023)}

\maketitle

\begin{abstract}
Metal inter-layer via (MIV) in Monolithic three-dimensional integrated circuits (M3D-IC) is used to connect inter-layer devices and provide power and clock signals across multiple layers.  The size of MIV is comparable to logic gates because of the significant reduction in substrate layers due to sequential integration. Despite MIV's small size, the impact of MIV on the performance of adjacent devices should be considered to implement IC designs in M3D-IC technology. In this work, we systematically study the changes in performance of transistors when they are placed near MIV to understand the effect of MIV on adjacent devices when MIV passes through the substrate. Simulation results suggest that the keep-out-zone (KOZ) for MIV should be considered to ensure the reliability of M3D-IC technology and this  KOZ is highly dependent on the M3D-IC process. In this paper, we show that the transistor placed near MIV considering the M1 metal pitch as the separation will have up to $68,668\times$ increase in leakage current, when the channel doping is $10^{15} cm^{-3}$, source/drain doping of $10^{18}cm^{-3}$ and substrate layer height of $100\ nm$. We also show that, this increase in leakage current can also be reduced significantly by having KOZ around MIV, which is dependent on the process.

%sequential integration of Monolithic 3D (M3D) integration, the vertical interconnect that connects between different device layers specifically metal inter-layer via (MIV) size has decreased considerably compared with the through silicon-via (TSV) in conventional 3D integration. Despite small size of MIV, the effect of MIV on the adjacent devices should be considered for proper implementation of Monolithic 3D integrated circuit (M3D-IC). In this work, we study the impact of MIV on the transistor placed near to the MIV considering the various process and design parameters of M3D-IC. Simulation results suggest that the transistor performance degradation is highly dependent on the M3D-IC process. Therefore, we need to have additional keep-out zone dependent on the process in order to design efficient M3D-ICs. From the simulation results, the leakage current of the transistor placed near MIV can be increased by up to (x63,000) for the assumed process parameters and this impact can be reduced by introducing process-aware keep-out zone for MIV in M3D-IC design.

\end{abstract}

%%
%% Keywords. The author(s) should pick words that accurately describe
%% the work being presented. Separate the keywords with commas.
%\keywords{ Monolithic 3D ICs, vertical integration, MIV}

%\section{Introduction}
\input{section1_intro}
\input{section2_koz}

%KOZ when parallel
\input{section2a_koz_parallel}

%KOZ when series
\input{section2b_koz_series}

%\section{Impact of Process Parameters in M3DIC}
\input{section3_process}

%\section{Design and process variations in detail}
\input{section5_table_contents_contd}

%\input{section5_table_old}
%\section{Conclusions and Future Work}
\input{section6_Conclusions}

\section*{Acknowledgements}
This work is supported by National Science Foundation under Award number -- 2105164.

\bibliographystyle{IEEEtran}
\bibliography{references}

\end{document}

%% file: section1_intro.tex
\section{Introduction}\label{sec:introduction}
Monolithic three-dimensional integrated circuit (M3D-IC) technology offers a promising solution to meet future computational needs. The need for M3D-IC technology arises since the conventional 2D-IC technology is limited by lithography and power constraints, whereas conventional three-dimensional integrated circuit (3D-IC) technology enabled by die stacking is limited by the through-silicon-via (TSV) size \cite{tida_efficacy_tvlsi,tida_multi_tier_resonant_tvlsi,tida_resonant_clocking_tvlsi,li2020improved}. The substrate layers in M3D-IC technology are realized by sequential integration. These layers are connected by metal inter-layer via (MIV), which are of the same size as logic gates \cite{shreepad_panth_miv}. This reduction in MIV size compared to the TSV has become possible by reduction in via height through sequential integration. However, to ensure the stability of the devices at the bottom layer, the top layers of M3D-IC should be processed below 500\textsuperscript{0} C \cite{thermal_budget_450,thermal_budget_450_2,thermal_budget_500,thermal_budget_500_2} which limits the M3D-IC process such as doping concentrations. 

MIVs are realized by metal surrounding oxide to provide electrical insulation to the substrate. However, this realization also forms metal-insulator-semiconductor (MIS) structure, and hence MIV can affect the substrate region around it \cite{madhava2020MWSCAS, uma_socc}. Therefore, the impact of MIV on the transistors around it should be studied to ensure the reliability of M3D-IC implementations. To the best of authors' knowledge, there are no commercial ICs with the defined M3D-IC process specifically doping concentration of top layers, substrate layer height and MIV thickness. Therefore, the process parameters for the IC design should be considered to understand the MIV impact on the devices around it. 

In this work, we systematically studied the impact of MIV on the adjacent devices and the simulation results suggest that the transistor placed near MIV considering the M1 metal pitch as the separation will have up to $68,668\times$ increase in leakage current specifically when the channel doping is $10^{15}\ cm^{-3}$, source/drain doping of $10^{18}\ cm^{-3}$ and substrate layer height of $100\ nm$. This impact can be reduced by increasing the keep-out-zone (KOZ) for MIV and this KOZ is highly dependent on the process parameters of M3D-IC technology. Therefore, process-aware KOZ is required to ensure the reliability of M3D-IC design. 

The organization of the rest of the paper is as follows: Section \ref{sec:M3DIc_process} discusses about the background and previous works on M3D-IC technology and the motivation of this work.  Section \ref{sec:effect_of_KOZ} demonstrates the impact of MIV on different transistor placement scenarios. Section \ref{sec:impact} investigates the process parameters influence on the transistor operation when MIV is placed at M1 metal pitch separation to the transistor. Section \ref{sec:process_variations} discusses the need for process-aware Keep-out-zone (KOZ) for the MIV, and KOZ values for different process parameters are presented. Section \ref{sec:future_directions} brings up the future directions for current work. The concluding remarks are given in Section \ref{sec:conclusions}.

%% file: section2_koz.tex
\section{Background and Motivation}\label{sec:M3DIc_process}

%In M3D-IC technology, the performance of bottom layer devices degrades significantly if the temperature is more than $500^{0}C$ for realizing the top layer devices \cite{thermal_budget_450, thermal_budget_450_2, thermal_budget_500, thermal_budget_500_2}. 

%Recent developments in M3D-IC have shown significant opportunities brought by the vertical integration. \cite{thermal_budget_450,thermal_budget_500,thermal_budget_500_2} have used low temperature process to ensure the stability of bottom layer devices. High temperature process used in dopant activation to create top layer devices, degrades the performance of the transistors in subsequent bottom layers. 
In M3D-IC technology, the top-layer devices should be processed sequentially under low-temperatures i.e., less than $500^{0}$C to ensure the quality of bottom-layer devices \cite{thermal_budget_450, thermal_budget_450_2, thermal_budget_500, thermal_budget_500_2}. With this sequential integration, the substrate layers are thinned down below 100nm thickness using layer transfer process \cite{silicon_thickness}.  With thin substrate and thin inter-layer dielectric (ILD), the metal inter-layer via (MIV) that connects devices at two different layers has the height of 100nm -- 500nm. This reduction in height helps us to reduce the MIV thickness significantly thus allowing high MIV density ($>10^{8}\ MIV/cm^{2}$) \cite{MIV_integration_density}. Recent works from CEA-LETI focus on sequential layer integration at lower temperatures and measured the device characteristics on both top and bottom layers with different integration techniques specifically Fully Depleted Silicon On Insulator (FDSOI) transistors \cite{3dvlsi_pdsoi, FDSOI_top_layer}. Several works on the M3D integration focuses on the partitioning, placement and routing in Monolithic 3D integration considering different top-layer device technology specifically bulk \cite{mono3D, Emre-Bulk} or FDSOI \cite{finfet_7nm_monolithic, FDSOI_top_layer}. Monolithic 3D-FPGA is designed using thin-film-transistors (TFT) in the lop layer \cite{TFT_top_layer}. \cite{RCAT_top_layer} have used the unique structure of recessed-channel-transistors in top layer to study its performance. Some M3D-IC based open libraries are available currently. \cite{mono3D,Emre-Bulk} use conventional CMOS technology to present 45nm library for M3D integration. \cite{finfet_7nm_monolithic} have studied the power benefit of M3D-IC using FinFET based 7nm technology node. Although several demonstrations of M3D-IC process are performed, there are no existing commercial industry standards \cite{Emre-Ind}. 

\begin{figure}[htp] 
    \centering
    \includegraphics[width=0.8\linewidth]{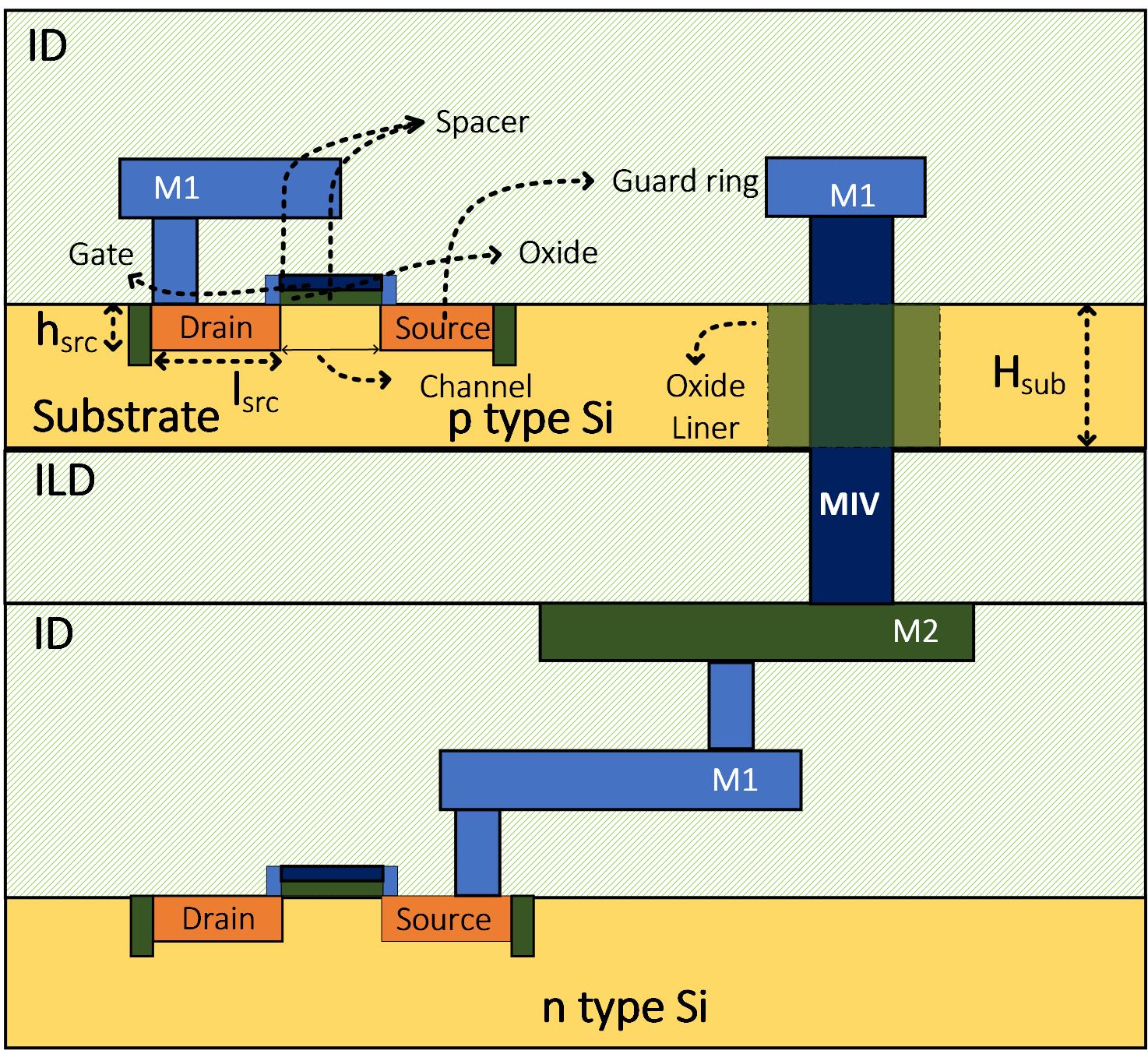}
    \caption{2-Layer M3D Process used in our work} \label{fig:miv_2layer_m3dprocess}
\end{figure}

In this work, we focus on the SOI devices on the top layer with thin substrate layer of up to 100 nm thickness as suggested by previous works \cite{silicon_thickness}. One of the major benefit of M3D-IC technology is to allow heterogeneous integration. This M3D-IC process will also consider the substrate biasing at the top-tier and hence tuning the threshold voltage of the transistor for analog and digital applications similar to the conventional MOSFET. % Also, this top-layer devices will not have significant cross-talk with the bottom-layer devices since top-layer device channel and active regions are realized at the top of the substrate layer \cite{SILICON THICKNESS 100NM suggestion works} compared with the ultra thin FDSOI devices. 
The two-layer M3D process assumed in this work is shown in Figure \ref{fig:miv_2layer_m3dprocess}.  With this consideration, the MIV passes through the substrate and essentially forms a metal-insulator-semiconductor (MIS) structure. Therefore, the impact of MIV on the adjacent devices should be studied since this MIS structure can interact through the substrate region around it. The rest of the paper will systematically study the impact of MIV on the adjacent devices considering the orientation of the channel with respect to the MIV.

\section{MIV Effect on Transistor Operation}\label{sec:effect_of_KOZ}

MIV with substrate around it forms a metal-insulator-semiconductor (MIS) structure, and essentially works like a MOS capacitor \cite{madhava2020MWSCAS,uma_socc}. If there is a p-type substrate region around MIV then it can be inverted (or form an n-type region) when carrying high voltage. This effect of MIV is shown in Figure \ref{fig:miv_channel_interactions}, where the substrate around MIV will become n-type region if a higher voltage is applied to the MIV because of the MIS structure. The extent of this inversion region depends on the voltage of MIV and the M3D-IC process. If there is an n-type transistor placed near MIV, then this effect should be considered since there is a possibility of forming resistive region between source and drain, or source/drain and MIV. %\redHL{ The inversion region around the MIV interacts with the channels of the devices around it resulting in changed current behavior. This is observed for all regions of operation of transistor including saturation and cut-off. As the shown in figure \ref{fig:miv_channel_interactions} when the distance increases from MIV placement, the inversion region is reduced, which arises a need to place devices at a minimum separation from MIV placement. However, the minimum separation depends on transistor performance. Therefore, we have to study its performance based on MIV placement to decide a minimum spacing for KOZ[Here need to add MIV and channel interactions]} 
Therefore, an n-type transistor placed near MIV can have significant impact on its characteristics. 

\iffalse
\begin{figure}[htp] 
\centering
\begin{subfigure}[b]{\linewidth}
\includegraphics[width=\linewidth]{Figures/layouts/miv_closer5.png}
\caption{}\label{subfig:miv_closer}
\end{subfigure}
\begin{subfigure}[b]{\linewidth}
\includegraphics[width=\linewidth]{Figures/layouts/miv_far.png}
\caption{}\label{subfig:miv_far}
\end{subfigure}
\caption{MIV placement near channel} \label{fig:miv_channel_interactions}
\end{figure} 
\fi

\begin{figure}[htp] 
    \centering
    \includegraphics[width=0.8\linewidth]{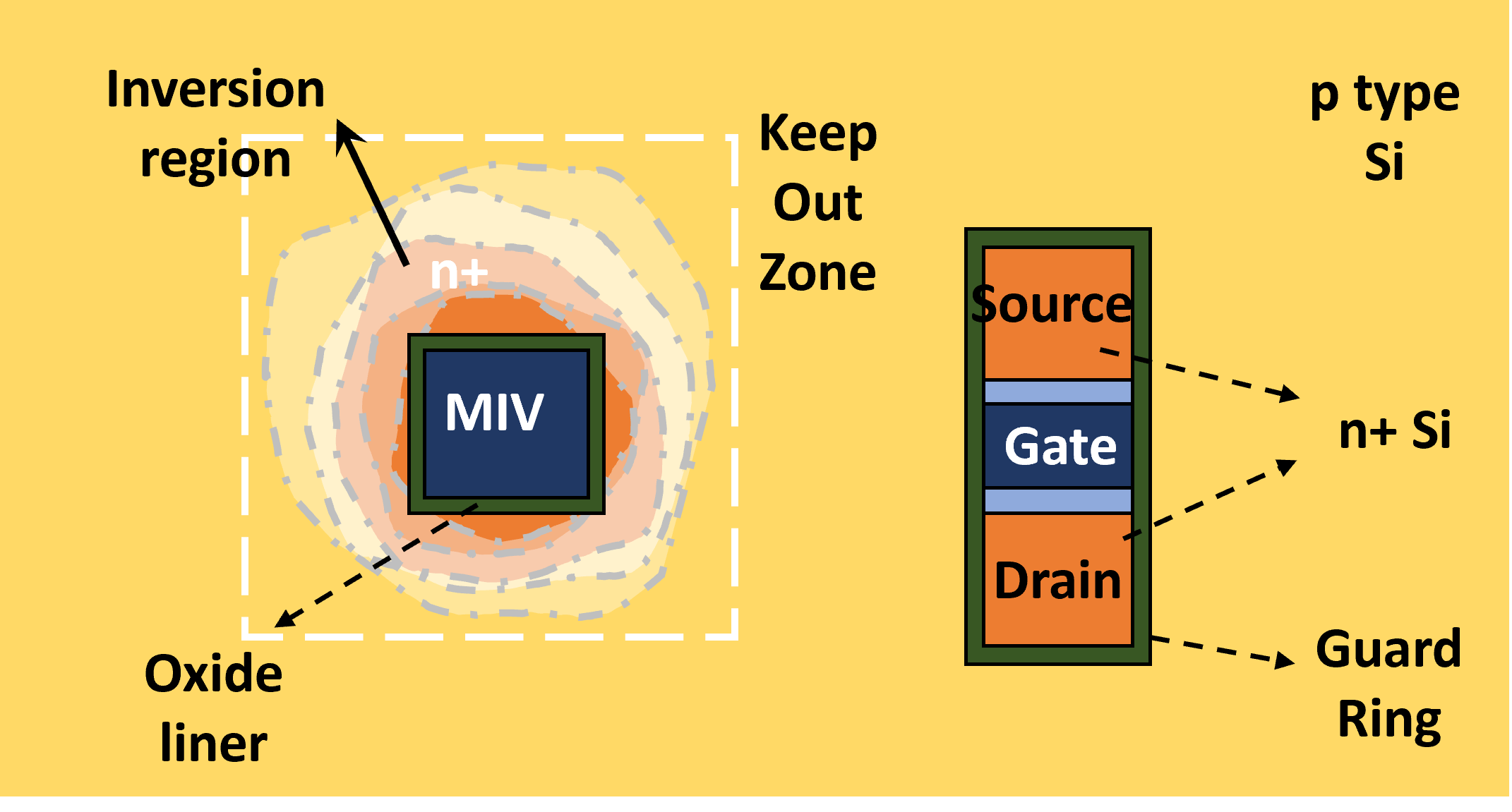}
    \caption{MIV interacting with adjacent device} \label{fig:miv_channel_interactions}
\end{figure}

In this section, we study the MIV affect on transistor operation by systematically investigating two scenarios of MIV placement with respect to the transistor and is as follows: \begin{enumerate}
    \item Vertical placement -- MIV is placed parallel to the transistor as shown in Figure \ref{fig:layout_parallel}.
    \item Horizontal placement -- MIV is placed beside the transistor as shown in Figure \ref{fig:layout_series}.
    
\end{enumerate}  These scenarios are modeled using Sentaurus Technology Computer Aided Design (TCAD) software. We used Boron (B) as p-type substrate material, where carrier behavior is modeled with Shockley-Read-Hall (SRH) recombination model and Fermi-based statistics.  The source/drain regions of the transistor is doped with Arsenic (As) (n-type) material using Gaussian profile concentration.  The substrate terminal is formed with a highly doped p+ region on the substrate to provide substrate biasing. Although, not shown in the figure, the substrate contact is placed near the transistor. 

The process parameters with their nominal values and range considered for this paper is given in Table \ref{tab:process_parameters}. We considered Copper (Cu) as the interconnect metal for MIV, Silicon (Si) the substrate material and Silicon dioxide (SiO\textsubscript{2}) as the liner material. $t_{miv}$ is the thickness of MIV, $t_{ox}$ is the thickness of oxide liner to provide insulation from the substrate, $H_{sub}$ is the height of the substrate layer through which the MIV passes through, $n_{sub}$ is the substrate doping concentration, and $n_{src}$ is the source and drain doping concentrations of the transistor. The nominal value of $t_{miv}$ is assumed to be 50 $nm$ as discussed in the previous works \cite{sklim_IEDM}. The nominal value of $t_{ox}$ is assumed to be $1\ nm$ considering the scaling between the TSV to MIV as discussed in \cite{sklim_14nm,TSV_liner}.  The length ($l_{src}$) and depth (not shown in figure) of the source/drain regions to implement the transistor is assumed to be $32\ nm$ and $7\ nm$ respectively. The width ($w$) of the transistor is assumed to be $32\ nm$. The length of the channel is assumed to be $14\ nm$, and the thickness of gate oxide is assumed to be 1 $nm$.  The thickness and depth of guard ring are assumed to be $7\ nm$ and $10\ nm$. The MIV pitch is assumed to be $100\ nm$  \cite{MIV_pitch}.

 We consider two performance metrics of the transistor to study the impact of MIV: \begin{enumerate}
    \item Maximum drain current ($I_{D,max}$), which is $I_D$ at $V_{GS} = 1\ V$ and $V_{DS} = 1\ V$.
    \item Maximum drain leak current ($I_{D,leak}$), which is $I_D$ at $V_{GS} = 0\ V$ and $V_{DS} = 1\ V$.
\end{enumerate}
We also assume that the voltage on the MIV ($V_{MIV}=1$) since it inverts the region around MIV (to n-type) due to MIS structure. 

In this section, we assume the nominal values for process parameters and the only variable considered is the placement of MIV with respect to transistor.

\begin{table}[htp]
\begin{center}
\caption{Process parameters with their range}\label{tab:process_parameters}
\begin{tabular}{|c|c|c|c|}
    \hline
    \textbf{Parameter} &  \textbf{Description } & \textbf{Value} & \textbf{Range}\\
    \hline
    t\textsubscript{miv} ($nm$) & MIV thickness  & $50$ & $20 \sim 100 $ \\ \hline
     t\textsubscript{ox} ($nm$) &  Liner thickness  & $1$  & $0.25 \sim 2 $\\ \hline
    H\textsubscript{sub} ($nm$) & Substrate height   & $100$  & $20 \sim 150 $ \\ \hline
    n\textsubscript{sub} ($cm^{-3}$) & Substrate doping  & $10^{17}$  & $ 10^{15} \sim 5\times10^{17} $ \\ \hline
    n\textsubscript{src} ($cm^{-3}$) & Source / Drain doping  & $10^{19}$  & $ 10^{18} \sim 10^{21} $ \\ \hline
   
\end{tabular}
\end{center}
\end{table}

%In this section we study the impact of MIV placement on the substrate and its effect on the transistor performance. Towards this, 2 orientations have been studied, 

 %to transistor, as shown as in figure \ref{fig:transistor_orientation}. In each orientation the distance has been varied based on a practical range of interest, and transistor performance is studied when MIV is located at a particular distance from transistor. In this study for the sake of simplicity we have assumed MIV passes through the substrate where only NMOS transistors are present. When the MIV passes through the substrate, to prevent it from interacting within the substrate, an insulating oxide liner is created around it. The parasitic around substrate and MIV in conjunction with oxide liner create a (Metal-Insulator-substrate) structure. When voltage is applied on MIV, it creates a resistive region around it affecting the transistor performance. The detailed effect of presence of MIV in the susbtrate is illustrated in the following sections.

%% file: section2a_koz_parallel.tex
 \subsection{Scenario 1 -- Vertical placement of MIV}

\begin{figure}[htp] 
\centering
\includegraphics[width=0.7\linewidth]{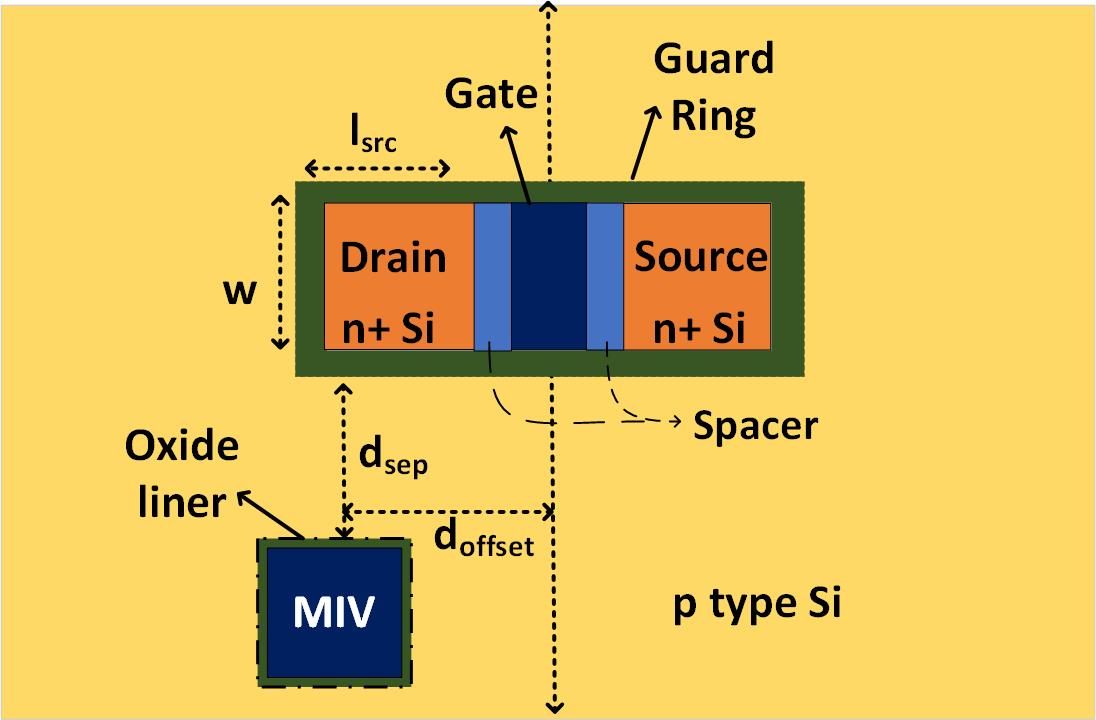}
\caption{MIV placement to transistor channel in vertical placement scenario (model not to scale)} \label{fig:layout_parallel}
\end{figure}

 In this scenario, MIV is placed in parallel to the transistor channel as shown in figure \ref{fig:layout_parallel}. In this subsection, we first focus on the effect of MIV on the transistor characteristics, when channel and MIV centers are aligned. Second, we study the impact of the offset distance from transistor channel to the center of MIV ($d_{offset}$) on the transistor performance. Finally, the transistor performance will be analyzed when the distance between the MIV and the transistor ($d_{sep}$) is varied.

\subsubsection{\textbf{MIV affect on transistor characteristics when channel center and MIV center are aligned}}

\begin{figure}[htp] 
\centering
\begin{subfigure}[b]{1.2\linewidth}
\includegraphics[width=1.0\linewidth]{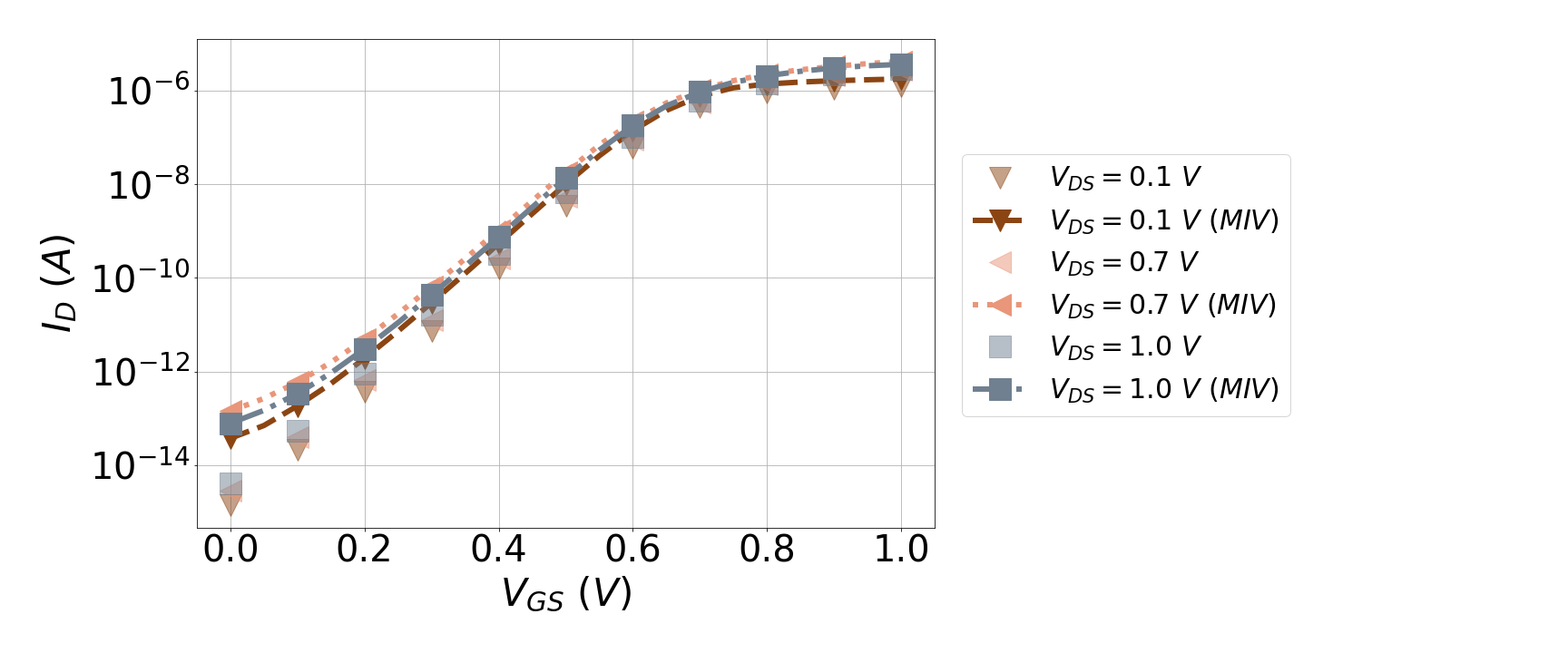}
\caption{$I_D$ v.s. $V_{GS}$}\label{subfig:IDVG_comparison}
\end{subfigure}
\begin{subfigure}[b]{1\linewidth}
\includegraphics[width=1.0\linewidth]{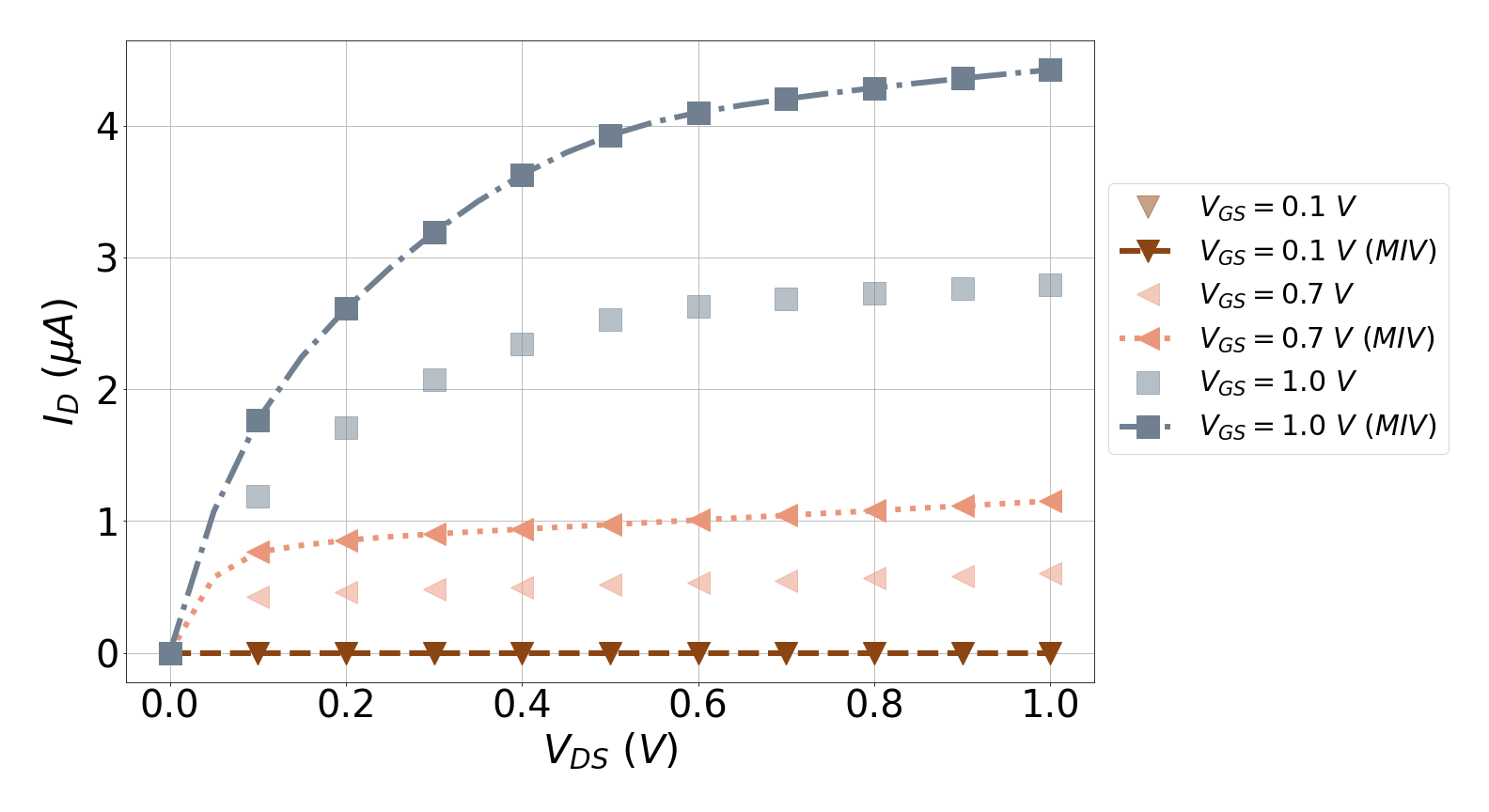}
\caption{$I_D$ v.s. $V_{DS}$}\label{subfig:IDVD_comparison}
\end{subfigure}
\caption{Nominal v.s. MIV effect on transistor characteristics} \label{fig:comparison_for_nominal_case}
\end{figure} 

 The impact of MIV on the transistor characteristics when $d_{sep} = 50\ nm$ and $d_{offset}=0$ is shown in Figure \ref{fig:comparison_for_nominal_case}. The drain current $I_D$ v.s. gate-source voltage $V_{GS}$ plots for different drain-source voltage $V_{DS}$ along with the ideal case where there is no MIV is shown in Figure \ref{subfig:IDVG_comparison}. Similarly, $I_D$ v.s. $V_{DS}$ for different $V_{GS}$ is shown in Figure \ref{subfig:IDVD_comparison}. From the figure, we see that the $I_{D,max}$ increases by up to $1.58\times$ and $I_{D,leak}$ increases by $70\times$. The increase of $I_{D,leak}$ by $70\times$ is a major concern since it will affect the power and thermal reliability of the IC.

\subsubsection{\textbf{d\textsubscript{offset} affect on transistor}}
\begin{figure}[htp] 
\centering
\includegraphics[width=0.8\linewidth]{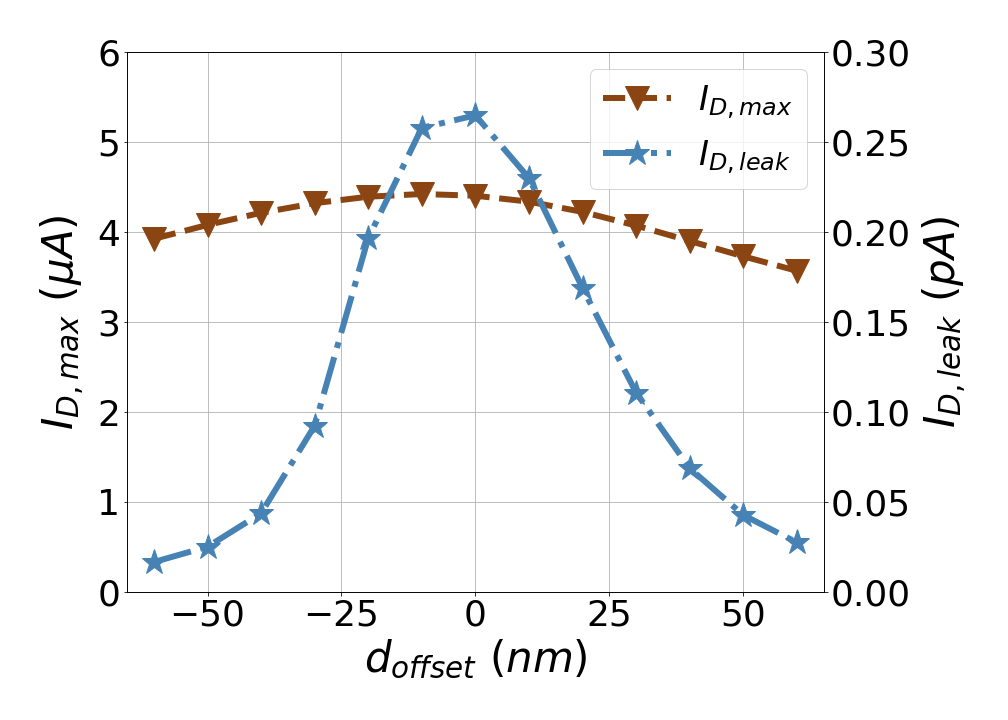}
\caption{$I_{D,max}$ and $I_{D,leak}$ v.s. $d_{offset}$ for $d_{sep}$ = $50\ nm$}\label{fig:offset_DC}
\end{figure}

 The $I_{D,max}$ and $I_{D,leak}$ v.s. $d_{offset}$ when $d_{sep} = 50\ nm$ is shown in Figure \ref{fig:offset_DC}, where the $I_{D,max}$ at $d_{offset}=50\ nm$ increased $1.34\times$ compared with the $I_{D,max}$ of transistor without MIV. The $I_{D,max}$ is maximum at $d_{offset} = 0\ nm$ and is increased by $1.58\times$ compared with transistor without MIV. Similarly $I_{D,leak}$ is maximum when $|d_{offset}|$ is low and is about $70\times$ compared with the transistor without MIV. This $I_{D,leak}$ is reduced to $6.6\times$ when $|d_{offset}|$ is high. Therefore, placing MIV such that the transistor channel and MIV centers are not aligned is a good practice for leakage reduction.

%% Check if the doffset is + or - for 46% case (34% also exists)

\subsubsection{\textbf{d\textsubscript{sep} affect on transistor}}
\begin{figure}[htp] 
\centering
\includegraphics[width=0.8\linewidth]{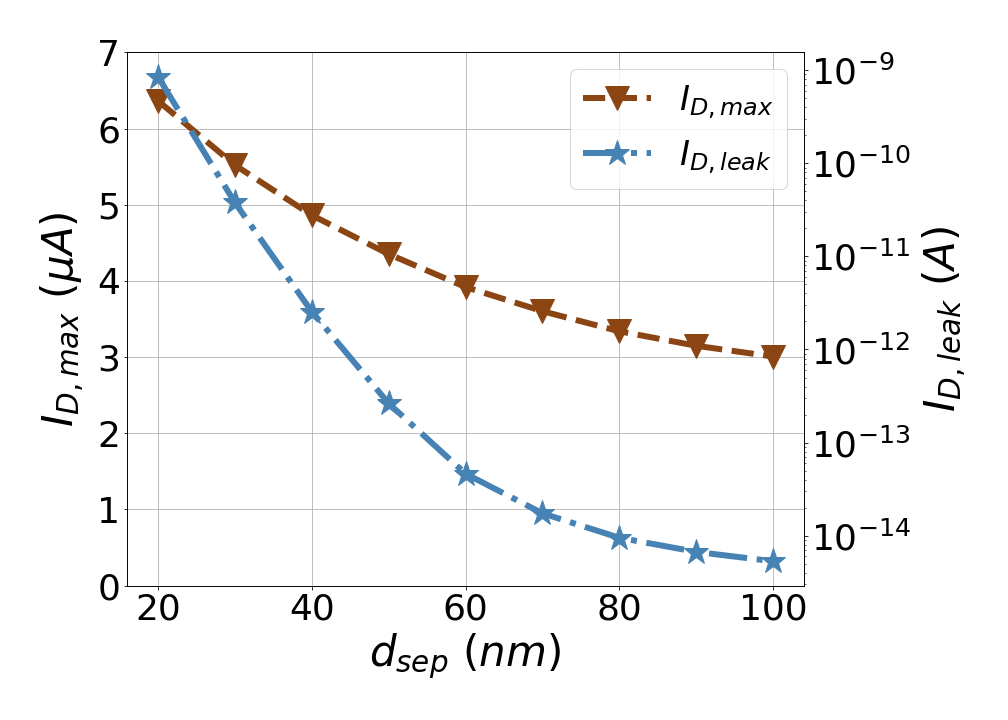}
\caption{$I_{D,max}$ and $I_{D,leak}$ v.s. $d_{sep}$ for $d_{offset}$ = $0\ nm$}\label{fig:parallel_DC}
\end{figure}

%We consider $I_{D,max}$ and $I_{D,leak}$ to investigate the impact of $d_{sep}$ on the transistor characteristics. 
The $I_{D,max}$ and $I_{D,leak}$ v.s. $d_{sep}$ at $d_{offset}=0$ is shown in Figure \ref{fig:parallel_DC}, where the $I_{D,max}$ and $I_{D,leak}$ reduces significantly with increase in $d_{sep}$. At $d_{sep} = 20\ nm$, $I_{D,max}$ increased to $2.2\times$ where as $I_{D,leak}$ increased to $225,400\times$ compared with the transistor characteristics without MIV. Also, please note the log axis for $I_{D,leak}$ in the Figure \ref{fig:parallel_DC}. At higher $d_{sep}$ i.e., at $100\ nm$, the $I_{D,max}$ increases by $1.08\times$ and $I_{D,leak}$ increases by $1.41\times$ compared with the transistor characteristics without MIV. Therefore, $d_{sep}$ has significant impact on the leakage of the transistor and should be considered as a design consideration for MIV placement to ensure proper M3D-IC realizations. \textit{Please note that, $d_{sep}$ is the distance between MIV and the transistor and can be considered as the keep-out-zone for MIV}.

% \redHL{offset case 1.1-16x for max current}.
% \redHL{1.58-122x for leakage current}.
 
%The distance of MIV from the transistor is another important factor that needs to be considered.

% \subsubsection{\underline{When offset is present}}
%  \redHL{Add the offset IDmax, IDLEAK plots here}

 %\subsubsection{\underline{When offset is absent}}
  %In this orientation the MIV is placed parallel  to the transistor channel with no deviation in the placement location. The effect of MIV separation on transistor performance metrics such as maximum drain current $I_{D,max}$ (current when $V_{GS}=1V$ and $V_{DS}=1V$) and $I_{leak,ratio}$ (ratio of leakage current when $V_{GS}=0V$ and $V_{DS}=1V$) is shown in figure \ref{fig:perf_study_parallel_to_channel}.

%\paragraph{Comparison with and without MIV}
%\redHL{Add IDVG and IVD plots here}

%\paragraph{transistor performance metrics vs separation of MIV}
%\redHL{add three subfigures here (Idleak, IDmax), (Idleak-transient),(Vth,SS) }

%% file: section2b_koz_series.tex
\subsection{Scenario 2 -- Horizontal placement of MIV}

\begin{figure}[htp]
\centering
\includegraphics[width=0.75\linewidth]{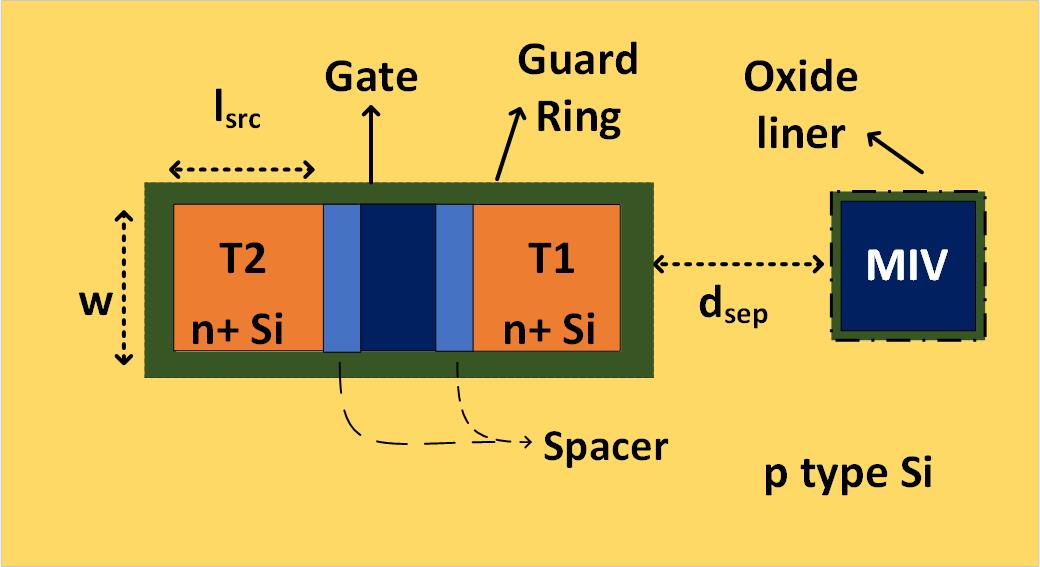}
\caption{MIV placement to transistor channel in horizontal placement scenario (model not to scale)} \label{fig:layout_series}
\end{figure}

In this scenario, MIV is placed horizontally or in series with the transistor active region where the centers of transistor channel and MIV are aligned as shown in Figure \ref{fig:layout_series}. The transistor is separated from the MIV by $d_{sep}$ distance where two cases are possible depending on the terminals $T_1$ and $T_2$: \ \begin{enumerate*}
     \item $T_1$ -- source and $T_2$ -- drain and,
     \item $T_1$ -- drain and $T_2$ -- source 
 \end{enumerate*}.
 
\begin{figure}[htp] 
\centering
\includegraphics[width=0.8\linewidth]{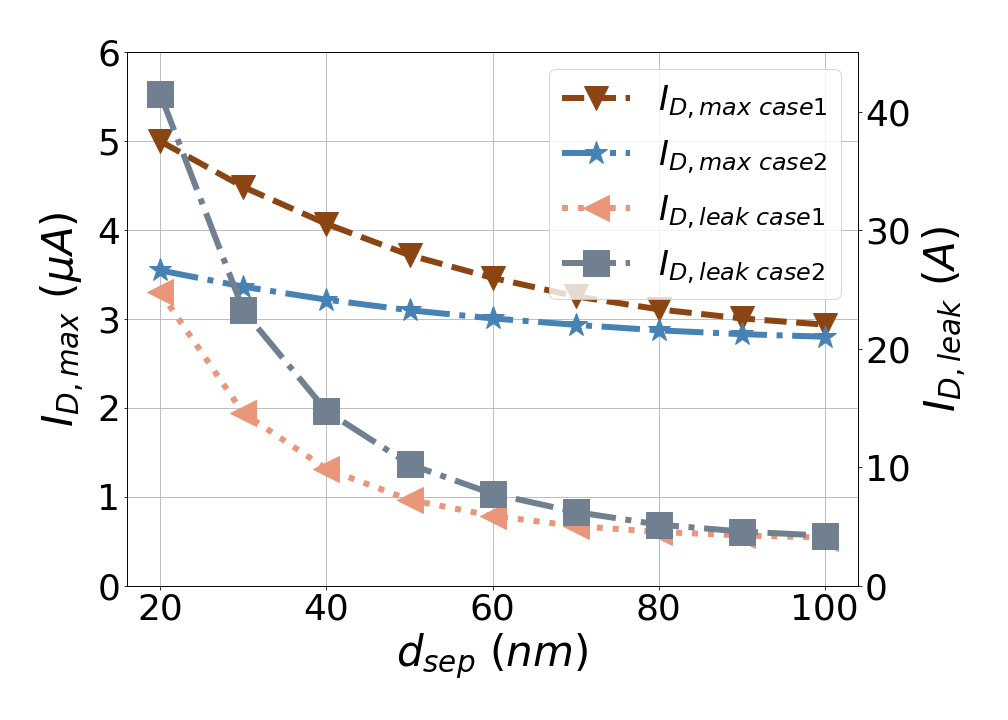}
\caption{$I_{D,max}$ and $I_{D,leak}$ v.s. $d_{sep}$}\label{subfig:series_performance_metrics_DC}
 \label{fig:series_performance}
\end{figure}

The $I_{D,max}$ and $I_{D,leak}$ v.s. $d_{sep}$ for the two cases i.e., case 1 -- $T_1$ as source and case 2 -- $T_1$ as drain is shown in Figure \ref{subfig:series_performance_metrics_DC}. The $I_{D,max}$ and $I_{D,leak}$ decreases with increase in $d_{sep}$. At $d_{sep}= 20\ nm$, the $I_{D,max}$
increased by $1.8\times$ and $1.3\times$ compared with the transistor without MIV for case 1 and case 2 respectively. $I_{D,leak}$ increased by $6.5\times$ and $11\times$ compared with the transistor without MIV at $d_{sep}=20\ nm$ for case 1 and case 2 respectively. At higher $d_{sep}$ i.e., at $100\ nm$, the $I_{D,max}$ increase only by $1.1\times$ and 1$\times$ for case 1 and case 2 respectively compared with the transistor without MIV. Similarly, $I_{D,leak}$ increase only by $1.1\times$ and $1.11\times$ for case 1 and case 2 respectively compared with the transistor without MIV. Therefore, $d_{sep}$ should be as higher as possible for scenario 2 to ensure $I_{D,leak}$ not to increase significantly. Although not discussed, the offset distance between MIV and transistor channel reduces both $I_{D,max}$ and $I_{D,leak}$ and should be as high as possible. 

\textbf{Observation 1:} Vertical placement of MIV shown in Figure \ref{fig:layout_parallel} has significant affect on transistor characteristics and require more keep-out distance compared with the horizontal placement of MIV shown in Figure \ref{fig:layout_series}.

However, eliminating the vertical placement scenario of MIV is not practically possible in order to obtain higher integration density. Therefore, in the rest of the paper, we consider only the vertical placement scenario of MIV and the same conclusions will also be valid for horizontal placement scenario.

%% file: section3_process.tex
\section{Impact of Process Parameters on Transistor Characteristics}\label{sec:impact}

One important consideration for M3D-IC design is the process parameters since the sequential integration should be achieved at low temperatures specifically below $500^\circ$C \cite{thermal_budget_500,thermal_budget_500_2}. Therefore, the impact of MIV on the transistor characteristics by varying process is an essential study to make practical design considerations since there is no defined process for M3D IC technology. In addition, with the rising demand for heterogeneous integration and mixed-signal IC designs, we believe that it is essential to investigate process parameter affect on transistor characteristics in M3D-IC. For clarity purposes, we consider only Scenario 1 i.e., vertical placement of MIV for this study. First, we systematically study the impact of process parameters on the transistor characteristics in the presence of MIV at $d_{sep}=50\ nm$ in M3D-IC using control variable method to change one parameter at a time. The nominal values of these parameters are given Table \ref{tab:process_parameters}. %Finally, we vary the process parameters that affect the transistor characteristics significantly to provide insight on practical considerations for MIV placement. 

%We consider two parameters: \begin{enumerate*}
%    \item Maximum drain current ($I_{D,max}$), which is $I_D$ at $V_{GS} = 1$ and $V_{DS} = 1$ and, 
%    \item Maximum drain leak current ($I_{D,leak}$), which is $I_D$ at $V_{GS} = 0$ and $V_{DS} = 1$
%\end{enumerate*}, to study the impact of process parameter on transistor characteristics. We assume that the voltage on the MIV ($V_{MIV}$=1) since it inverts the region around MIV (to n-type) due to MIS structure.

\subsection{\textbf{MIV thickness (t\textsubscript{miv})}}
\begin{figure}[!htp] 
\centering
\includegraphics[width=0.8\linewidth]{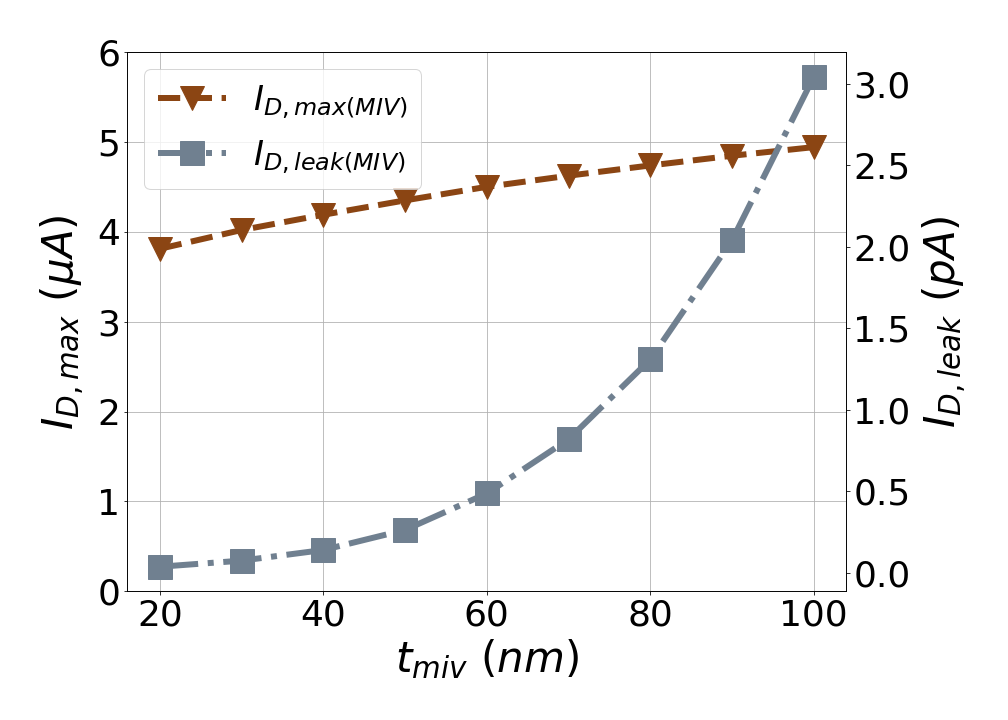}
\caption{$I_{D,max}$ and $I_{D,leak}$ v.s. $t_{miv}$} \label{fig:vary_tmiv}
\end{figure}

$I_{D,max}$ and $I_{D,leak}$ v.s. $t_{miv}$ is shown in Figure \ref{fig:vary_tmiv}. The nominal $I_{D,max}$ and $I_{D,leak}$ obtained for the transistor without MIV is $2.79\ \mu A$ and $3.78\ fA$ respectively. \textit{Note that the transistor characteristics without MIV does not change with $t_{miv}$.}  From the figure, we can see that the $I_{D,max}$ increases almost linearly from $1.4\times$, when $t_{miv} = 20nm$ to $1.8\times$, when $t_{miv}=100\ nm$ compared with nominal $I_{D,max}$. Also, the $I_{D,leak}$ increases  from $10\times$, when $t_{miv} = 20\ nm$ to $805\times$, when $t_{miv} = 100\ nm$ compared with the nominal $I_{D,leak}$. Therefore, the MIV affect is more prominent on the $I_{D,leak}$, and this effect should be considered while placing MIV near transistors. From previous observations, we know that $I_{D,leak}$ will reduce with $d_{sep}$, and hence the increase in $t_{miv}$ require more $d_{sep}$ (or KOZ) from the transistor to ensure lower $I_{D,leak}$.

\subsection{\textbf{MIV liner thickness (t\textsubscript{ox})}}

\begin{figure}[htp] 
\centering
\includegraphics[width=0.8\linewidth]{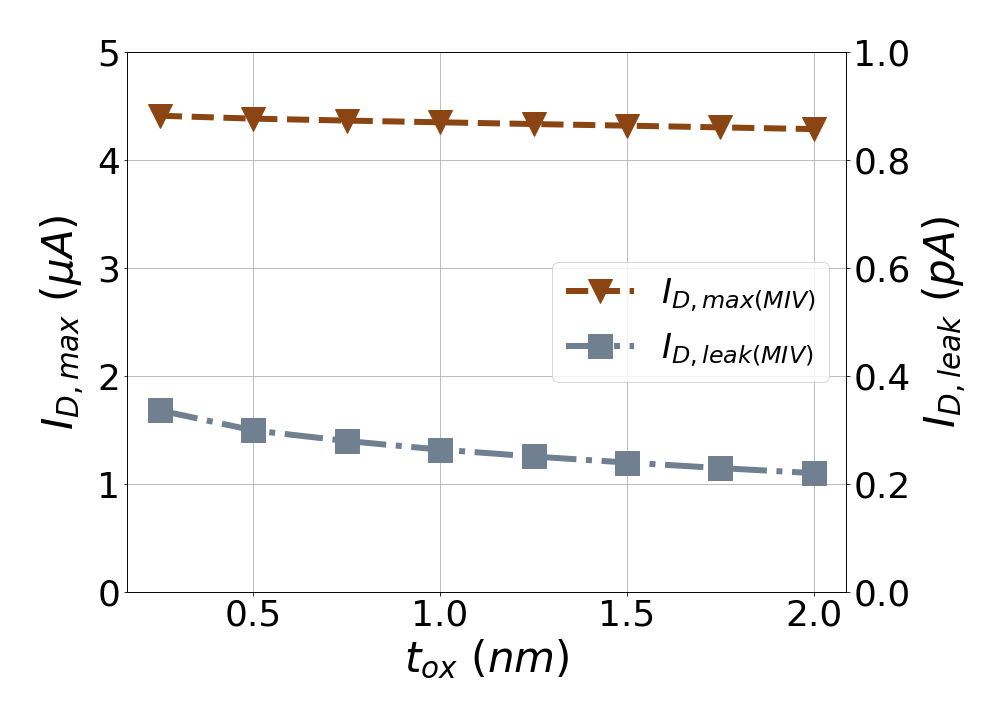}
\caption{$I_{D,max}$ and $I_{D,leak}$ v.s. $t_{ox}$} \label{fig:vary_tox}
\end{figure}

$I_{D,max}$ and $I_{D,leak}$ v.s. $t_{ox}$ is shown in Figure \ref{fig:vary_tox}. The nominal $I_{D,max}$ and $I_{D,leak}$ obtained for the transistor without MIV is $2.79\ \mu A$ and $3.78\ fA$ respectively. \textit{Note that the transistor characteristics without MIV does not change with $t_{ox}$}.  From the figure, we can see that the $I_{D,max}$ almost remains constant at around $1.58\times$ increase compared with nominal $I_{D,max}$. $I_{D,leak}$ decreases with increase in $t_{ox}$ where it is $89\times$, when $t_{ox}=0.25\ nm$ to $58\times$, when $t_{ox}= 2\ nm$ compared with the nominal $I_{D,leak}$. Therefore, higher $t_{ox}$ is desired for reducing MIV impact on leakage current.

\subsection{\textbf{Height of Substrate (H\textsubscript{sub})}}

\begin{figure}[htp] 
\centering
\includegraphics[width=0.8\linewidth]{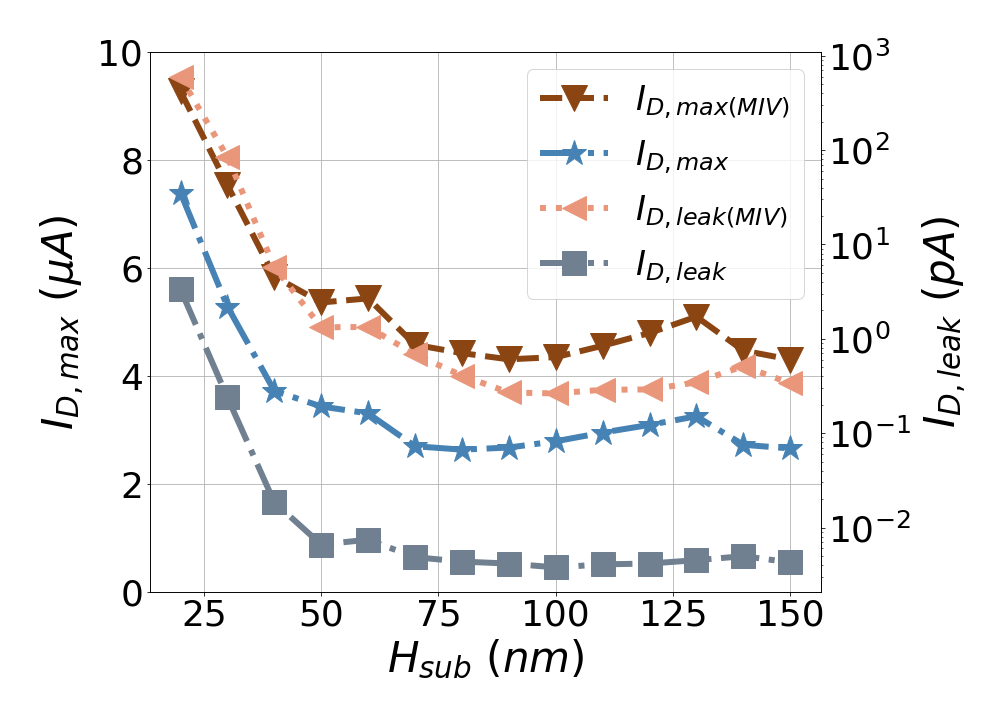}
\caption{$I_{D,max}$ and $I_{D,leak}$ v.s. $H_{sub}$} \label{fig:vary_Hsub}
\end{figure}

$I_{D,max}$ and $I_{D,leak}$ v.s. $H_{sub}$ obtained for transistor with MIV and without MIV is shown in Figure \ref{fig:vary_Hsub}. In this case, the transistor characteristics without MIV will also change with $H_{sub}$ and therefore the $I_{D,max}$ and $I_{D,leak}$ of transistor without MIV is also included in Figure \ref{fig:vary_Hsub}, where the plots labeled as $I_{D,max}$ and $I_{D,leak}$ correspond to the transistor without MIV case and the plots labeled as $I_{D,max}(MIV)$ and $I_{D,leak}(MIV)$ corresponds to the transistor with MIV case. From the figure, the presence of MIV near the transistor increases both $I_{D,max}$ and $I_{D,leak}$ compared with the transistor without MIV presence. From the figure, we see that $I_{D,max}$ increases by up to $1.7\times$ in MIV presence at nominal separation compared with the case without MIV. Similarly, $I_{D,leak}$ increases by up to $353\times$. \textit{One important thing to consider is that the height of substrate also affects the minimum thickness of MIV due to the change in aspect ratio of MIV and hence we cannot increase the substrate height significantly.}

\subsection{\textbf{Substrate Doping (n\textsubscript{sub})}}

\begin{figure}[htp] 
\centering
\includegraphics[width=0.8\linewidth]{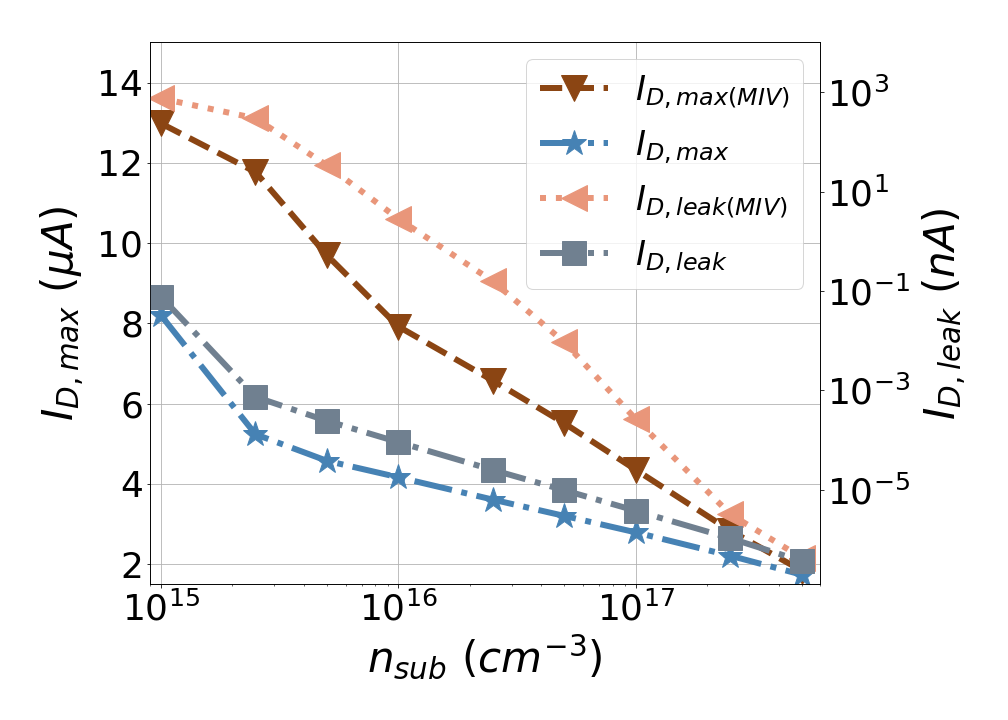}
\caption{$I_{D,max}$ and $I_{D,leak}$ v.s. $n_{sub}$} \label{fig:vary_nsub}
\end{figure}

$I_{D,max}$ and $I_{D,leak}$ v.s. $n_{sub}$ characteristics for transistor with MIV and without MIV is shown in Figure \ref{fig:vary_nsub}. From the figure, we see that as $n_{sub}$ increases, the $I_{leak}$ decreases for both transistor in presence of MIV and transistor without MIV cases. However, the presence of MIV increases both $I_{D,max}$ and $I_{D,leak}$ of the transistor compared with the transistor without MIV. We also found that $I_{D,max}$ and $I_{D,leak}$ increases by up to $2.25\times$ and  $403,100\times$ respectively with MIV presence. Also, the nominal $n_{src}$ is $10^{19}\ cm^{-3}$  and therefore if the $n_{src}/{n_{sub}}$ ratio  is higher, the leakage will be very high because of higher reverse saturation current at drain and substrate boundary. In addition, we know that the depletion region between the drain and substrate will increase with the decrease of substrate doping $n_{sub}$ \cite{depletionwidth} and therefore the impact of MIV on $I_{D,leak}$ is higher at the lower $n_{sub}$.

\subsection{\textbf{Active Doping (n\textsubscript{src})}}

\begin{figure}[htp] 
\centering
\includegraphics[width=0.8\linewidth]{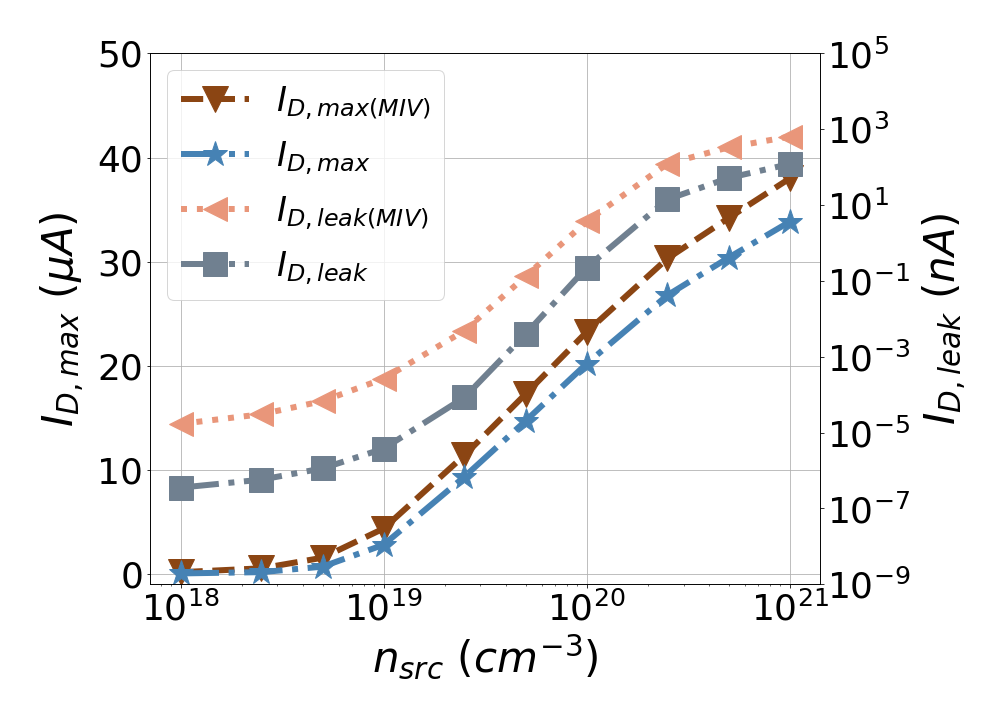}
\caption{$I_{D,max}$ and $I_{D,leak}$ v.s. $n_{src}$} \label{fig:vary_nsrc}
\end{figure}

$I_{D,max}$ and $I_{D,leak}$ v.s. $n_{src}$ is shown in Figure \ref{fig:vary_nsrc}. From the figure, we see that the $I_{D,max}$ increases in the linear scale where as the $I_{D,leak}$ increases exponentially for both transistor with MIV and without MIV cases. As $n_{src}$ increases up to $10^{19}\ cm^{-3}$, presence of MIV resulted in increased leakages ranging from $48\times$ to $70\times$. We also observed that the $I_{D,max}$ increases by up to $4.23\times$ and $I_{D,leak}$ increases by up to $70\times$ for transistor with MIV presence compared with the transistor without MIV. We also observed that $I_{D,leak}$ increases significantly with the increase of $n_{src}$ and, therefore the $n_{src}/n_{sub}$ ratio should not be very high.

\textbf{Observation 2:} $n_{sub}$ has significant impact on the leakage current of the transistor, when the other process parameters are assumed to be nominal values shown in Table \ref{tab:process_parameters} and MIV is placed at $d_{sep} = 50\ nm$. 

%% file: section5_table_contents_contd.tex
\input{section5_table_contents}

\section{Keep-out-Zone for MIV in M3D-IC process}\label{sec:process_variations}

In this section, we consider three process parameters, specifically substrate height $H_{sub}$, substrate doping $n_{sub}$ and source/drain doping $n_{src}$ to study the impact of MIV at the assumed process for realizing the transistor at M3D-IC technology. We assume that $n_{src}/n_{sub}$ to be $100$ and $1000$ and varied the $n_{src}$ from $ 10^{18}\ cm^{-3}$ to $ 10^{21}\ cm^{-3}$. The substrate height $H_{sub}$ is also varied from $25\ nm$ to $100\ nm$. For these cases, we assume the thickness of MIV to be $50\ nm$ and the liner thickness to be $1\ nm$. The impact of MIV placed at $50\ nm$ away from the transistor of $32\ nm$ width on the transistor characteristics is shown in Table \ref{tab:process_variations}. From the table, we can see that at higher $n_{src}$ (i.e., last three rows colored in blue), the leakage is not significantly increased with MIV presence compared with the transistor leakage without MIV. At lower $n_{sub}$, the MIV presence has significant impact on the leakage where the $I_{leak}$ increased more than $100\times$ compared with the transistor leakage without MIV (i.e., top rows colored in red) and is increased by up to $68,668\times$. Therefore, the M1 metal pitch of $100\ nm$ ($d_{sep} = 50\ nm$) is not sufficient to ensure the reliability of M3D-IC design. From Table \ref{tab:process_KOZ}, we can say that keep-out-zone (KOZ or minimum $d_{sep}$) between MIV and the transistor is needed to ensure reliability of the M3D-IC realization, and is highly dependent on the $n_{sub}$, $n_{src}$ and $H_{sub}$ of the transistor.

\begin{table}[htp]
    \centering
   \caption{KOZ or minimum $d_{sep}$ (in nm) for different process parameters of M3D-IC}\label{tab:process_KOZ}
   \scalebox{0.9}{
    \begin{NiceTabular}{|c|c|c|c|c|c|}
         \hline 
        $n_{src}$  &  $n_{sub}$ & \multicolumn{4}{c|}{$H_{sub}(nm)$} \\ \cline{3-6} 
          
         $(cm^{-3})$ &  $(cm^{-3})$ & {$25$}& {$50$} & {$75$}& {$100$} \\ 
          \hline
          \multirow{2}{*}{$10^{18}$}
          & $10^{15}$
                % hsub=25nm % hsub=50nm % hsub=75nm % hsub=100nm
                & 400 & 450 & 500 & 500\\ \cline{2-6}
        & $10^{16}$
                % hsub=25nm % hsub=50nm % hsub=75nm % hsub=100nm
                & 200 & 200 & 200 & 200 \\ \hline

          \multirow{2}{*}{$10^{19}$}
          & $10^{16}$
                % hsub=25nm % hsub=50nm % hsub=75nm % hsub=100nm
                & 200 & 250 & 200 & 200 \\ \cline{2-6}
        & $10^{17}$
                % hsub=25nm % hsub=50nm % hsub=75nm % hsub=100nm
                & 100 & 100 & 100 & 100\\ \hline

          \multirow{2}{*}{$10^{20}$}
          & $10^{17}$
                % hsub=25nm % hsub=50nm % hsub=75nm % hsub=100nm
                & 50 & 100 & 100 & 100 \\ \cline{2-6}
        & $10^{18}$
                % hsub=25nm % hsub=50nm % hsub=75nm % hsub=100nm
                & 50 & 50 & 50 & 50 \\ \hline
        \multirow{2}{*}{$10^{21}$}
          & $10^{18}$
                % hsub=25nm % hsub=50nm % hsub=75nm % hsub=100nm
                & 50  & 50 & 50 & 50 \\ \cline{2-6}
        & $10^{19}$
                % hsub=25nm % hsub=50nm % hsub=75nm % hsub=100nm
                & 50  & 50 & 50 & 50 \\ \hline
       
    \end{NiceTabular} 
    }

\end{table}

\textit{\textbf{Note:} KOZ is defined as the minimum spacing around MIV where no other active devices should be formed. $d_{sep}$ is the distance between MIV and the transistor placed near by. Therefore, the minimum $d_{sep}$ required and KOZ are same.}

Table \ref{tab:process_KOZ} shows the KOZ in nm for the different process parameters of the transistors where we assumed that the KOZ increases in steps of $50\ nm$. The KOZ value is obtained when  the $I_{leak}$ of the transistor placed near MIV is less than $10\times$ compared to the transistor without MIV. From the table, we can see that transistor process-aware KOZ for MIV is essential for proper operation of the M3D-IC design where the KOZ value can range between $50\ nm$ to $500\ nm$ for the assumed variations in the process parameters specifically $n_{src}$, $n_{sub}$ and $H_{sub}$. Therefore, at floorplanning and placement stage of M3D-IC designs, we need to consider these KOZ considerations, and is dependent on the nearby transistor specifications. For example, assume a transistor near MIV has the process parameters as $H_{sub} = 100\ nm$, $n_{sub} =  10^{17}\ cm^{-3}$  and $n_{src} =  10^{19}\ cm^{-3}$ then  the KOZ for MIV should be at least $100\ nm$.

%% file: section5_table_contents.tex
\begin{table*}[htp]
    \centering
   \caption{Maximum drain current $I_{D,max}$ and Leakage current $I_{D,leak}$ for different process parameters of M3D-IC}\label{tab:process_variations}
   \scalebox{0.69}{
    \begin{NiceTabular}{|c|c|c|c|c|c|c|c|c|c|c|}
         \hline
          & $n_{src} $ & $n_{sub} $ & \multicolumn{2}{c|}{$H_{sub}=25 nm$}& \multicolumn{2}{c|}{$H_{sub}=50 nm$} & \multicolumn{2}{c|}{$H_{sub}=75 nm$}& \multicolumn{2}{c|}{$H_{sub}=100 nm$}  \\ \cline{4-11} 
        & $(cm^{-3})$ &  $(cm^{-3})$ & \multicolumn{1}{c|}{$I_{max} \ (\mu  A)$} & \multicolumn{1}{c|}{$I_{leak} \ (A)$} 
        & \multicolumn{1}{c|}{$I_{max} \ (\mu  A)$} & \multicolumn{1}{c|}{$I_{leak} \ (A)$}
        & \multicolumn{1}{c|}{$I_{max} \ (\mu  A)$} & \multicolumn{1}{c|}{$I_{leak} \ (A)$}
        & \multicolumn{1}{c|}{$I_{max} \ (\mu  A)$} & \multicolumn{1}{c|}{$I_{leak} \ (A)$} \\
        \hline 
        \multirow{8}{*}{\rotatebox{90}{$d_{sep}=50nm$}}
        & \multirow{2}{*}{$10^{18}$} 
    %   nsub % Imax_with_MIV % Ileak_withMIV
        & $10^{15}$
                % hsub=25nm
               
                &  \cellcolor{red_t}  $4.60\ (\times1.30)$ & \cellcolor{red_t}  $1.65\times10^{-07}\ (\times171)$ 
                % hsub=50nm
                & \cellcolor{red_t}  $4.48\ (\times1.64)$ & \cellcolor{red_t}  $2.63\times10^{-07}\ (\times882)$
                % hsub=75nm
                & \cellcolor{red_t} $4.66\ (\times2.01)$ & \cellcolor{red_t}  $3.04\times10^{-07}\ (\times8090)$\cellcolor{red_t} 
                % hsub=100nm
                & \cellcolor{red_t} $4.55\ (\times2.53)$ & \cellcolor{red_t}  $3.50\times10^{-07}\ (\times68668)$ \\  \cline{3-11}
        
        & & $10^{16}$
                % hsub=25nm
                & \cellcolor{red_t} $2.89\ (\times2.00)$ & \cellcolor{red_t}  $5.34\times10^{-09}\ (\times1865)$ 
                % hsub=50nm
                & \cellcolor{red_t} $1.71\ (\times3.30)$ & \cellcolor{red_t}  $1.01\times10^{-09}\ (\times13906)$
                % hsub=75nm
                & \cellcolor{red_t} $1.45\ (\times5.90)$ & \cellcolor{red_t}  $4.56\times10^{-10}\ (\times31241)$
                % hsub=100nm
                & \cellcolor{red_t} $1.30\ (\times6.53)$ & \cellcolor{red_t}  $3.10\times10^{-10}\ (\times44983)$  \\ \cline{4-11}

        \cline{2-3}
        & \multirow{2}{*}{$10^{19}$} 
        & $10^{16}$   
                % hsub=25nm
                & \cellcolor{red_t} $11.89\ (\times1.29)$ & \cellcolor{red_t}  $8.93\times10^{-08}\ (\times442)$ 
                % hsub=50nm
                & \cellcolor{red_t}  $9.34\ (\times1.56)$ & \cellcolor{red_t}   $1.25\times10^{-08}\ (\times6988)$
                % hsub=75nm
                & \cellcolor{red_t} $8.23\ (\times1.84)$ & \cellcolor{red_t}  $5.00\times10^{-09}\ (\times18709)$
                % hsub=100nm
                & \cellcolor{red_t}  $7.92\ (\times1.90)$ & \cellcolor{red_t}  $2.83\times10^{-09}\ (\times31175)$   \\ \cline{3-11}

        & & $10^{17}$
                % hsub=25nm
                & \cellcolor{red_t}   $8.57\ (\times1.33)$ & \cellcolor{red_t}  $1.78\times10^{-10}\ (\times244)$ 
                % hsub=50nm
                & \cellcolor{red_t}   $5.37\ (\times1.56)$ & \cellcolor{red_t}    $1.33\times10^{-12}\ (\times201)$
                % hsub=75nm
                & \cellcolor{grey_t}  $4.51\ (\times1.59)$ & \cellcolor{grey_t}  $4.61\times10^{-13}\ (\times99)$
                % hsub=100nm
                & \cellcolor{grey_t}   $4.35\ (\times1.56)$ & \cellcolor{grey_t}  $2.64\times10^{-13}\ (\times70)$ \\ \cline{4-11}

        \cline{2-3}
        & \multirow{2}{*}{$10^{20}$} 
        & $10^{17}$   
                % hsub=25nm
                & \cellcolor{blue_t} $30.60\ (\times1.12)$ & \cellcolor{blue_t}  $8.23\times10^{-07}\ (\times7)$ 
                % hsub=50nm
                & \cellcolor{grey_t}   $24.92\ (\times1.15)$ & \cellcolor{grey_t}  $2.15\times10^{-08}\ (\times26)$
                % hsub=75nm
                &  \cellcolor{grey_t}  $23.51\ (\times1.16)$ & \cellcolor{grey_t}  $8.51\times10^{-09}\ (\times26)$
                % hsub=100nm
                &  \cellcolor{grey_t}  $23.24\ (\times1.15)$ & \cellcolor{grey_t}  $3.74\times10^{-09}\ (\times17)$  \\ \cline{3-11}
        
        & & $10^{18}$
                % hsub=25nm
                & \cellcolor{blue_t} $17.89\ (\times1.01)$ & \cellcolor{blue_t} $3.36\times10^{-11}\ (\times1.13)$ 
                % hsub=50nm
                &  \cellcolor{blue_t} $14.92\ (\times1.00)$ & \cellcolor{blue_t} $2.63\times10^{-12}\ (\times1.09)$
                % hsub=75nm
                & \cellcolor{blue_t} $14.81\ (\times1.01)$ & \cellcolor{blue_t} $2.35\times10^{-12}\ (\times1.21)$
                % hsub=100nm
                & \cellcolor{blue_t} $15.09\ (\times1.01)$ & \cellcolor{blue_t} $2.41\times10^{-12}\ (\times1.07)$ \\ \cline{4-11}
        
        \cline{2-3}
        
        & \multirow{2}{*}{$10^{21}$} 
        & $10^{18}$  
                % hsub=25nm
                & \cellcolor{blue_t} $29.21\ (\times1.02)$ & \cellcolor{blue_t} $3.60\times10^{-08}\ (\times1.16)$ 
                % hsub=50nm
                & \cellcolor{blue_t} $24.61\ (\times1.02)$ & \cellcolor{blue_t} $2.37\times10^{-09}\ (\times1.20)$
                % hsub=75nm
                & \cellcolor{blue_t} $24.36\ (\times1.01)$ & \cellcolor{blue_t} $2.25\times10^{-09}\ (\times1.10)$
                % hsub=100nm
                & \cellcolor{blue_t} $24.93\ (\times1.01)$ & \cellcolor{blue_t} $3.60\times10^{-09}\ (\times1.09)$  \\ \cline{3-11}
        
        & & $10^{19}$ 
                % hsub=25nm
                & \cellcolor{blue_t} $4.01\ (\times1.03)$ & \cellcolor{blue_t} $3.79\times10^{-15}\ (\times1.16)$ 
                % hsub=50nm
                &  \cellcolor{blue_t}
                $4.01\ (\times1.02)$ & \cellcolor{blue_t} $4.45\times10^{-15}\ (\times1.13)$
                % hsub=75nm
                & \cellcolor{blue_t} $3.47\ (\times1.00)$ & \cellcolor{blue_t} $3.01\times10^{-15}\ (\times1.03)$
                % hsub=100nm
                & \cellcolor{blue_t} $4.32\ (\times1.00)$ & \cellcolor{blue_t} $6.39\times10^{-15}\ (\times1.00)$   \\ \cline{4-11}
        
        \hline 
    \end{NiceTabular} 
    }
\end{table*}

%% file: section6_Conclusions.tex
\section{Future Directions}\label{sec:future_directions}

In this work, we focused on the thin silicon substrate of 25nm -- 100 nm thickness for top-layer in M3D-IC technology. Our future work will also focus on the effect of MIV when multiple transistors are placed around it and the process-aware KOZ of MIV-based design optimization for M3D-IC circuits. However, there are also demonstrations on the FDSOI device with ultra-thin channel in the range of 6nm -- 10nm \cite{FDSOI_top_layer,VT_stability_FDSOI,thermal_budget_450,thermal_budget_500,3dvlsi_pdsoi,analog_issue_m3DIC}. For this top-layer realization, the back-gate bias for top-layer FDSOI devices should be investigated for coupling reduction from the bottom-layer devices and interconnects. Although the MIV does not pass through the substrate for top-layer FDSOI devices, the device will be adjacent to the MIV depending on the MIV minimum distance. The channel with the MIV will also form the MIS structure but with thick oxide defined by the separation between MIV and adjacent device. Therefore, the MIV impact on the device characteristics should also be considered for reliable top-layer devices since MIV can potentially turn on the channel and increase leakage due to the capacitive coupling. Also, FinFET based top-layer devices are also realized in M3D-IC technology \cite{finfet_7nm_monolithic}. Similar study for the effect of FinFET device characteristics with MIV adjacent to it should also be considered for the reliable M3D-IC implementations.

Capacitive coupling between MIV and the substrate is considered to realize MIV-devices thus reducing the MIV area overhead in \cite{madhava2020MWSCAS,uma_socc}. Similar study considering this MIS structure for ultra-thin FDSOI devices can be considered. In addition, the placement and routing considerations of power delivery networks (PDN) and clock distribution networks (CDN) considering MIVs to route between layers should be studied.

\section{Conclusions}\label{sec:conclusions}

In this work, we have discussed the impact of MIV on the surrounding substrate region and on the transistor placed near MIV. We have performed a systematic study on the effect of MIV on the characteristics of the transistor when an MIV is placed near the transistor at different orientations. We then demonstrated that the process parameters specifically substrate doping and source/drain doping of the transistor have a significant effect on the leakage current of the transistor. Finally, we studied the minimum KOZ requirement for the M3D-IC process where we assumed that the substrate doping, source doping and substrate height as parameters and obtained KOZ for each possible process parameter with the assumed M3D-IC process.